\documentstyle[12pt]{article}
\begin{document}
\vspace*{2cm}
\begin{center}
\begin{large}
{\bf Correlation energy of  fractional-quantum-Hall-effect\\
systems of composite fermions}
\end{large}
\end{center}

\vspace{1cm}
\begin{center}
Piotr Sitko and Lucjan Jacak
\vspace{0.3cm}

  Institute of Physics,
Technical University of Wroc\l{}aw,\\ Wyb. Wyspia\'{n}skiego
 27, 50-370 Wroc\l{}aw, Poland.
\end{center}

 \vspace{1cm}
\begin{center}

Abstract

\end{center}
We find the RPA correlation energy and collective modes of
the systems of composite fermions
corresponding to the fillings of $1/3$, $1/5$, $2/5$ in the
fractional quantum Hall effect. It is verified that transmutations to
composite fermions do not change the ground-state energy.

\newpage
\noindent{\bf 1.  Introduction}
\vspace{0.5cm}

The fractional quantum Hall effect (FQHE) is originally explained within the
idea of the Laughlin wave function \cite{Laughlin}. An alternative approach
was proposed by Jain \cite{Jain,Greiter} who noted that in
two-dimensional (2D) systems
the antisymmetricity of the many-particle wave
function  is not an unambiguous criterion of quantum statistics.
The antisymmetricity is held by a class of quantum-statistics
particles called
composite fermions \cite{Jain}. An interchange of two
composite fermions produces a
phase factor of $e^{i(2p+1)\pi}$ (p - an integer number). Note that
this is also the case of the Laughlin wave function.

Let us consider 2D electron gas in the external magnetic field.
Statistics transmutations to composite fermions
are given by  Chern-Simons
gauge field which is equivalent to attaching an even number ($2p$)
of flux quanta
to each electron \cite{Lopez1,Lopez2}:
\begin{equation}
H=\frac{1}{2m}\sum_{i=1}^{N}({\bf P}_{i}+\frac{e}{c}{\bf
A}_{i}+\frac{e}{c}{\bf
A}_{i}^{ex})^{2}
\end{equation}
where
\begin{equation}
{\bf A}_{i}=\frac{-2p \hbar c}{e}	\hat{\bf z}\times \sum_{j\neq i}
\frac{({\bf r}_{i}-{\bf r}_{j})}{|{\bf r}_{i}-{\bf r}_{j}|^{2}}\;\; ,
\end{equation}
$\hat{\bf z}$ - a perpendicular to the plane unit vector.
Replacing the sum of poin fluxes by an average flux one finds
the average statistical field
$B^{s}=-2p\frac{hc}{e}\rho$ which reduces
the effective field acting on electrons to
$B^{ef}=B^{ex}+B^{s}$. We predict an analog of the quantum Hall
effect when, effectively, $n$ Landau levels a completely filled, i.e.
$B^{ef}=\frac{1}{n} \frac{hc}{e}\rho$. Hence,
$B^{ex}=\frac{2pn+1}{n} \frac{hc}{e}\rho$, which means that from the
point of view of the external magnetic field the lowest Landau level
is filled in the fraction $\nu=\frac{n}{2pn+1}$. The RPA calculations
confirms that the system (1) exibits the fractional quantum Hall
effect \cite{Lopez1,Sitko1}.

Considering the Hartree-Fock approximation of the system	(1) one
finds the ground-state energy
\cite{Sitko1,Sitko2}:
\begin{equation}
\label{HFenergia}
\frac{<H>}{N}=\frac{1}{2}E_{F}(1+3p^{2}+\frac{2p}{n}-\frac{p^{2}}{n}).
\end{equation}
where $E_{F}=\frac{2\pi\hbar^{2}\rho}{m}$ - the Fermi
energy of the 2D electron gas. The energy of electrons in the external
magnetic field
$B^{ex}=\frac{2pn+1}{n} \frac{hc}{e}\rho$ is
 $\frac{1}{2}NE_{F}(2p+\frac{1}{n})$. Comparing with the result
(\ref{HFenergia})
one finds the cost of the transmutation to composite fermions:
\begin{equation}
\frac{\Delta E}{N}=\frac{1}{2}E_{F}[2p^{2}+(p-1)^{2}(1-\frac{1}{n})]
\end{equation}
which is always greater than zero and increases rapidly with $p$.

One of the crucial ideas supporting	the existance of new realisations
of quantum statistics in
2D systems is the braid group argument \cite{Chen}.
However,  composite fermions
have  the same braid group representation
as fermions. Hence, it is reasonable to expect
transmutations to composite fermions not to change the ground-state
energy.
In this paper we verify the
result (\ref{HFenergia}) by calculating the RPA correlation energy for
filling fractions $1/3$, $1/5$ and $2/5$.	We find also dispersion
relations of collective modes at these fractions.

\vspace{0.5cm}
\noindent{\bf 2.  RPA correlation energy}
\vspace{0.5cm}

The Hamiltonian $H$ can be separated into two parts:
$H=H_{0}+H_{1}$
where\\ $H_{0}=\frac{1}{2m}\sum_{i}({\bf P}_{i}+{\bf A}^{ef}_{i})^{2}$
is  treated as the
unperturbed term and
$H_{1}$ is the {\it interaction} Hamiltonian	 \cite{Chen,Dai}.
In this paper we
assume that $B^{ef}=\nabla\times {\bf
A}^{ef}=B^{ex}+B^{s}=\frac{1}{n} \frac{hc}{e}\rho$ and in the ground
state one has $n$ completely filled Landau levels.
Let us define the current density:
\begin{equation}
\bf j \rm ({\bf r})=\frac{1}{2m}\sum_{j}\left\{{\bf P}_{j}
+\frac{e}{c}{\bf A}_{j}^{ef},\delta({\bf r} -{\bf r}
_{j})\right\}
\end{equation}
where braces denote an anticommutator,
  {\bf j} is the vector part
of $j^{\mu}$ with $\mu=0,x,y$. We define
$j^{0}$ as density fluctuations:
$j^{0}=\sum_{j}\delta
({\bf r} -{\bf r} _{j})-\rho$.

It can be shown that the correlation energy (if omitting three-body
contributions) is given by \cite{Hanna,Szewczenko}
\begin{equation}
E_{c}^{RPA}	= -\frac{1}{2}\hbar L^{2}\int\frac{d{\bf q}}{(2\pi)^{2}}
\int_{0}^{\infty}\frac{d\omega}{\pi}\int_{0}^{1}\frac{d\lambda}{\lambda}
{\rm Im\; tr} (\lambda V(q))[D_{\lambda}^{RPA}({\bf q},\omega)-D_{0}({\bf
q},\omega )]
\end{equation}
where $D_{\lambda}^{RPA}$ is the correlation function of
effective field currents:
\begin{equation}
D_{\mu\nu}^{RPA}({\bf r} t,{\bf r'} t')=
-\frac{i}{\hbar}<T[j^{\mu}({\bf r} t),j^{\nu}({\bf r'} t')]>
\end{equation}
given within the random-phase approximation (with the coupling
constant $\lambda$):
\begin{equation}
D_{\lambda}^{RPA}
({\bf q} ,\omega)=[I-\lambda D_{0}
({\bf q} ,\omega)V({\bf q})]^{-1}D_{0}
({\bf q} ,\omega).
\end{equation}
The interaction matrix $V$ is obtained from
the Hamiltonian $H_{1}$. We choose ${\bf q}=q\hat{\bf x}$ and the Coulomb
gauge
 which reduce the problem to $2\times 2$ $D^{RPA}_{\mu\nu}$
 matrix ($\mu=0,y$) \cite{Halperin}.
Taking  $\omega_{c}^{ef}=\frac{eB^{ef}}{cm}$ and
$a_{0}^{ef}=\sqrt{\frac{\hbar c}{eB^{ef}}}$
to be
frequency and length units,
respectively one finds \cite{Sitko1,Halperin}:
\begin{equation}
V({\bf q} )=\frac{4p\pi}{q^{2}}\left(
\begin{array}{cc}
2pn&-iq\\
iq&0
\end{array}\right) .
\end{equation}
As in the case of anyons  \cite{Jacak} we have
\begin{equation}
D_{0}
({\bf q} ,\omega)=\frac{n}{2\pi}\left(
\begin{array}{cc}
q^{2}\Sigma_{0}&-iq\Sigma_{1}\\
iq\Sigma_{1}&\Sigma_{2}
\end{array}\right)
\end{equation}
where
\begin{displaymath}
\Sigma_{j}=\frac{e^{-x}}{n}\sum_{m=n}^{\infty}\sum_{l=0}^{n-1}
\frac{m-l}{(\omega)^{2}-(m-l-i\eta)^{2}}
\frac{l!}{m!}x^{m-l-1}[L_{l}^{m-l}(x)]^{2-j}
\end{displaymath}
\begin{equation}
\times [(m-l-x)L_{l}^{m-l}(x)+2x\frac{dL_{l}^{m-l}(x)}{dx}]^{j}
\end{equation}
and $x=\frac{q^{2}}{2}$ ($L_{l}^{m}$ -- Laguerre polynomials).
Then one obtains:
\begin{equation}
D^{RPA}
({\bf q} ,\omega)=\frac{n}{2\pi det}\left(
\begin{array}{cc}
q^{2}\Sigma_{0}&-iq\Sigma_{s}\\
iq\Sigma_{s}&\Sigma_{p}
\end{array}\right)
\end{equation}
where
 $det=det(I-D^{0}V)=(1-2pn\Sigma_{1})^{2}-(2pn)^{2}\Sigma_{0}(1+\Sigma_{2})$,
 $\Sigma_{s}=\Sigma_{1}-2pn\Sigma_{1}^{2}+2pn\Sigma_{0}\Sigma_{2}$,\\
 $\Sigma_{p}=(2pn)^{2}\Sigma_{1}^{2}+\Sigma_{2}-(2pn)^{2}\Sigma_{0}
 \Sigma_{2}$.\\
Collective modes are determined by the poles of the correlation
function $D^{RPA}$ \cite{Chen}.
In plotting Fig. 1-3 we have used simpler relation finding
 zeros of the determinant $det$ \cite{Dai}.

In units of $\hbar\omega_{c}^{ef}$ the correlation energy can be
expressed as follows \cite{Hanna}:
\begin{equation}
E_{c}^{RPA}	= \frac{N}{2n}\int_{0}^{\infty} q dq
\int_{0}^{\infty}\frac{d\omega}{\pi}
{\rm Im} (\ln{det}+{\rm tr}(V(q)D_{0}))
\end{equation}
which is equal to
\begin{equation}
E_{c}^{RPA}	= \frac{N}{2n}\int_{0}^{\infty} q dq
\int_{0}^{\infty}\frac{d\omega}{\pi}
{\rm Im} (\ln{det}+2pn(2pn\Sigma_{0}+2\Sigma_{1})).
\end{equation}
Following the results of Hanna and Fetter \cite{Hanna} we can write:
\begin{equation}
\int_{0}^{\infty}dx\int_{0}^{\infty}\frac{d\omega}{\pi}
{\rm Im}\Sigma_{0}(x,\omega)=-\frac{1}{2}\sum_{m=1}^{\infty}\frac{1}{m}
+\frac{1}{2}(S_{n}-1)
\end{equation}
($S_{n}=\sum_{j=1}^{n}\frac{1}{j}$)	and
\begin{equation}
\int_{0}^{\infty}dx\int_{0}^{\infty}\frac{d\omega}{\pi}
{\rm Im}\Sigma_{1}(x,\omega)=0.
\end{equation}
In the following section we will calculate the remainig integral
numerically using dispersion relations of collective modes.

\vspace{0.5cm}
\noindent{\bf 3.  The results for $n=1$ and $n=2$}
\vspace{0.5cm}

Let us consider first  the case of the one filled Landau level ($n=1$), then
double poles in the determinant $det$, generally appearing in the expression
$\Sigma_{1}^{2}-\Sigma_{0}\Sigma_{2}$,
cancel out and we have an infinite set of modes with shortwavelength
behaviour like $\omega_{m}(q\rightarrow\infty)=m$ -- Fig.1-2.
One can see that at the frequency of $2pn+1$	(in units of
$\omega_{c}^{ef}$ and thus at $\omega_{c}^{ex}=\frac{eB^{ex}}{cm}$)
the pole at $q\rightarrow 0$ is
degenerated which was first predicted by Lopez and Fradkin
\cite{Lopez2}.
It can
be shown that \cite{Hanna}:
\begin{equation}
\int_{0}^{\infty}\frac{d\omega}{\pi}{\rm Im} \ln{det}=\sum_{m=1}^{\infty}
(\omega_{m}-m)=\sum_{m=1}^{\infty}\Delta\omega_{m}
\end{equation}
and then
\begin{equation}
\frac{E_{c}^{RPA}}{N}	= \frac{1}{2}\sum_{m=1}^{\infty}
(\int \Delta\omega_{m}(x)dx -(2p)^{2}\frac{1}{2m}).
\end{equation}
The integrals have been calculated numerically using $k$-point
Gauss-Laquerre integration \cite{ksiazka}. It was verified that the
summation over $m$ converges well and the sums have been truncated at
$2k$ terms.    The results for filing fractions $1/3$, $1/5$  are given
in Table I and show that large Hartre-Fock contributions are
canceled out.

\vspace{1cm}
\begin{center}
\begin{tabular}{|c|c|c|c|}
\hline\hline
k&1/3&1/5&2/5\\  \hline
10&-1.032&-5.292&-2.248\\
15&-1.032&-5.282&-2.248\\
20&-1.032&-5.279&-2.247\\ \hline
Approximated value &-1.032 &-5.274 &-2.246 \\ \hline
$\frac{\Delta E}{N}$ in units of $E_{F}$&-0.032 &-1.274 &-0.123\\
\hline\hline
\end{tabular}
\end{center}

\vspace{0.3cm}
\noindent
Table I. RPA correlation energies for $\nu = 1/3$, $1/5$, $2/5$ (in
respective units of $\hbar\omega_{c}^{ef}$).
To obtain approximated value we have used a functional
dependence $A+Bk^{-2}$.
The cost of the transmutation to composite
fermions $\frac{\Delta E}{N}$ combines  Hartree-Fock energy and  RPA
correlation energy.

\vspace{1cm}
In the case of two occupied Landau levels ($n=2$) the doubles poles
in the determinant $det$ are present. In contrast to the
result of  Ref.\cite{Lopez2} every root higher than first is splitted
into two \cite{Hanna} (for $m>1$,
$\omega_{m}^{-}(q\rightarrow\infty)=m=\omega_{m}^{+}(q\rightarrow\infty)$
-- Fig. 3).
Again at $\omega_{c}^{ex}$
one finds at $q=0$
different  splitting, however, into three
roots. One has
\begin{equation}
\int_{0}^{\infty}\frac{d\omega}{\pi}{\rm Im}
\ln{det}=\Delta\omega_{1}+\sum_{m=2}^{\infty}
(\Delta\omega_{m}^{-}+\Delta\omega_{m}^{+})
\end{equation}
and the correlation energy is then given by
\begin{equation}
\frac{E_{c}^{RPA}}{N}	= \frac{1}{4}(\int \Delta\omega_{1}(x)dx
-8p^{2})
+\frac{1}{4}\sum_{m=2}^{\infty}
[\int (\Delta\omega_{m}^{-}(x)+\Delta\omega_{m}^{+}(x))dx
-(4p)^{2}\frac{1}{2m}]+p^{2}.
\end{equation}
The result for the filling of $2/5$ is given in Table I, as in the case
of $1/3$	the cost of the transmutation to composite fermions
is close to
zero.

\vspace{0.5cm}
\noindent{\bf 4.  Conclusions}
\vspace{0.5cm}

The FQHE systems of composite fermions are considered within the
random-phase approximation.
In agreement with the results of Lopez and Fradkin \cite{Lopez2} the
double degeneracy	of the collective mode at $\omega_{c}^{ex}$
(at $q=0$) for
one filled Landau level is found -- Fig.1-2.
However, in contrast to Ref. \cite{Lopez2}, for two filled Landau
levels we have the additional splitting into two roots
of each mode ($m\neq 1$) \cite{Hanna}. At $\omega_{c}^{ex}$ one finds
at $q\rightarrow 0$ the
splitting into three roots -- Fig. 3.

The RPA correlation energies at the fractions $1/3$, $1/5$, $2/5$
are obtained
verifying the Hartree-Fock results \cite{Sitko1}. For $p=1$ (the minimal
transmutation, $\nu=1/3$, $2/5$)
it is shown that the cost of the transmutation to composite fermions
is close to zero. However, for $p=2$ ($\nu=1/5$) the cost becomes
significantly
negative. It is suspected that such discrepancy comes from omitting
three-body contributions in the RPA calculations \cite{Hanna}. For  $p>1$ the
logarithmic interaction in the potential matrix $V(q)$ dominates
other contributions which affects the RPA result. Nevertheless, the
obtained results strongly suggest the cost of transmutations to composite
fermions to be zero.

\vspace{0.5cm}
\noindent{\bf Acknowledgement}
\vspace{0.5cm}

We would like to thank prof. John Quinn for helpful comments. We are
also grateful to Witold Prusinowski for help in
programming.

\end{document}